\documentclass[a4paper]{article}

\usepackage{amssymb,amsmath,amsthm}
\usepackage{epsfig,graphicx}
\usepackage{hyperref}
\usepackage{setspace}

\newtheorem{thm}{Theorem}
\newtheorem{prop}[thm]{Proposition}
\newtheorem{cor}[thm]{Corollary}
\newtheorem{obs}[thm]{Observation}
\newtheorem{lem}[thm]{Lemma}
\newtheorem{claim}[thm]{Claim}
\newtheorem{defn}[thm]{Definition}

\newcommand{\eat}[1]{}

\newcommand{\jodist}[1]{J_{#1}}
\newcommand{\nstates}{k}

\newcommand{\D}{\mathcal{D}}

\newcommand{\PP}{\mathcal{P}}
\newcommand{\EE}{{\mathbb E}}

\newcommand{\Q}{\mathcal{Q}}
\renewcommand{\S}{\mathcal{S}}

\newcommand{\cP}{\mathcal{P}}
\newcommand{\Co}{\mathit{Co}}
\newcommand{\tstar}{T_{\star}}
\newcommand{\ttwo}{T_{1}}
\newcommand{\tthree}{T_{2}}
\newcommand{\tfour}{T_{3}}
\newcommand{\sens}{\ensuremath{L}}
\newcommand{\cotstar}{$\Co(\tstar)$}
\newcommand{\cotwoisect}{$\Co(\ttwo) \cap \Co(\tthree)$}
\newcommand{\cottwo}{$\Co(\ttwo)$}
\newcommand{\cotwounion}{$\Co(\ttwo,\tthree)$}
\newcommand{\counion}{$\Co(\ttwo,\tthree,\tfour)$}
\newcommand{\pvect}{\underline{p}}
\newcommand{\qvect}{\underline{q}}
\newcommand{\onevect}{\underline{1}}
\newcommand{\eset}{\emptyset}
\newcommand{\ip}[2]{\langle#1,#2\rangle}
\newcommand{\polymake}{\texttt{polymake}}
\newcommand{\leaves}{\ell}
\newcommand{\mixarr}{\rightarrowtail}
\newcommand{\nomixarr}{\,\not\!\mixarr}
\newcommand{\splits}{\Sigma}
\newcommand{\restr}[2]{#1|_{#2}}
\newcommand{\ujoin}[2]{#1\hbox{---}\,#2}
\newcommand{\vect}[1]{\underline{#1}}
\newcommand{\Dstar}{\mathcal{D}^\star}
\newcommand{\Mike}[1]{\textbf{**Mike**}}

\author{Frederick A. Matsen,\thanks{Biomathematics Research Centre,
University of Canterbury, New Zealand. email: ematsen@gmail.com,
m.steel@math.canterbury.ac.nz.
Supported by the Allan Wilson Centre for Molecular Ecology and Evolution.} \
Elchanan Mossel,\thanks{
Statistics, UC Berkeley, USA. email: mossel@stat.berkeley.edu. Supported by a Sloan fellowship in Mathematics, NSF awards DMS 0528488 and DMS 0548249 (CAREER) and by ONR grant N0014-07-1-05-06} \
and Mike Steel$^\ast$
}

\title{Mixed-up trees: the structure of phylogenetic mixtures}

\begin{document}

\maketitle

\begin{spacing}{1}

\begin{abstract}
  In this paper we apply new geometric and combinatorial methods to
  the study of phylogenetic mixtures. The focus of the geometric
  approach is to describe the geometry of phylogenetic mixture
  distributions
  for the two state random cluster model, which is a generalization of
  the two state symmetric (CFN) model. In particular, we show that the set of mixture
  distributions forms a convex polytope and we calculate its dimension;
  corollaries include a simple criterion for when a mixture of
  branch lengths on the star tree can mimic the site pattern frequency
  vector of a resolved quartet tree. Furthermore, by computing volumes of polytopes
  we can clarify how ``common'' non-identifiable mixtures
  are under the CFN model.  We also present a new combinatorial result
  which extends any identifiability result for a specific pair of trees
  of size six to arbitrary pairs of trees. Next we present a positive
  result showing identifiability of rates-across-sites models. Finally,
  we answer a question raised in a previous paper concerning ``mixed
  branch repulsion'' on trees larger than quartet trees under the CFN
  model.
\end{abstract}

\begin{center}


\end{center}

\hspace{1cm}

Molecular phylogenetic inference methods reconstruct evolutionary
history from sequence data. Many years of research have shown that if
data evolves according to a single process under certain assumptions
then the underlying tree can be found given sequence data of
sufficient length. For an introduction to this literature see
\cite{felsenstein} or \cite{semple-steel}.

However, it is known that molecular evolution varies according to
position,
 even within a single gene \cite{pmid8752001}. Between genes even more
heterogeneity is observed \cite{pmid10830951}, though it is not
unusual for researchers to concatenate data from different genes for
inference \cite{rokas}. This poses a different challenge for
theoretical phylogenetics: is it possible to reconstruct
the tree from data generated by a combination of different processes?

This question is formalized as follows. The raw data for most
phylogenetic inference techniques is site-pattern frequency vectors, i.e.
normalized counts of how often certain data patterns occur. If
multiple data sets are combined, the corresponding site-pattern frequency
vectors are combined according to a weighted average.
In statistical terminology, this is called a ``mixture model.'' 
In the phylogenetic setting, there are various means of generating
a site-pattern frequency vector given a tree with edge parameters, for
example the expected frequency vector under a mutation model. 
\begin{defn}
  Assume some way of generating site-pattern frequency data from trees
  and edge parameters, i.e. a map $\psi$ from pairs
  $(T_i, \vect{\xi_i})$ to site pattern frequency vectors.
  We define a \emph{phylogenetic mixture} (on $h$ classes) to be any
  vector of the form
  \begin{equation}
    \label{eq:mixture}
    \sum_{i=1}^{h} \alpha_i \psi(T_i, \vect{\xi_i})
  \end{equation}
  where for each $i$, $\alpha_i > 0$ and $\sum_i \alpha_i = 1$. When
  all of the $T_i$ are the same, we call the phylogenetic mixture a
  \emph{phylogenetic mixture on a tree}.
\end{defn}
The formal version of our question is now ``given a phylogenetic
mixture (\ref{eq:mixture}) can we infer the trees $T_i$ and the edge
parameters $\vect{\xi_i}$?''

The answer to this question is certainly ``not always.''
In 1994 Steel~et.~al. \cite{pmid8790461} presented the first
``non-identifiable'' examples, i.e. phylogenetic mixtures on a
tree such that the underlying tree cannot be inferred from the data. More recently,
\v{S}tefankovi\v{c} and Vigoda \cite{stefankovic-vigoda} were the
first to explicitly construct such examples.
Even more recently, Matsen and Steel \cite{matsen-steel07}
showed the stronger statement that a phylogenetic mixture on one tree can ``mimic''
(i.e. give the same site-pattern frequency vector as) an unmixed
process on a tree of another topology.

This raises several questions, some of which are answered in this
paper for the two state models and some
generalizations. First, now that we know these non-identifiable examples
exist, is there some way of describing exactly which site-pattern
frequency vectors correspond to non-identifiable mixtures? Below we note that the set of mixture distributions on a
tree of a given topology forms a convex polytope with an simple description
(Proposition~\ref{prop:ST}); thus the non-identifiable patterns (being
a finite intersection of polytopes) 
form a convex polytope as well. Now, computing dimensions shows that
a ``random'' site-pattern frequency vector has a non-zero probability
of being non-identifiable, which raises the question of the relative volumes of
a given tree polytope and the non-identifiable polytopes. This question is
answered by computer calculations for the quartet case in
Table~\ref{tab:volumes}. We also show that surprisingly well-resolved
trees sit inside the phylogenetic mixture polytope for the star tree
(Proposition~\ref{prop:whichtreesinstar}). This same proposition
implies that the internal edge of a quartet tree must be long compared
to the pendant edges if the corresponding site-pattern frequency
vector is to be identifiable.

The second main section focuses on identifiability results for
mixtures of two trees under various assumptions. These results
partially ``bookend'' the non-identifiability results of
\cite{matsen-steel07,stefankovic-vigoda}. The first emphasis for this
work is combinatorial, answering the question (Theorem~\ref{disent}) ``if we know all of the splits
associated to the restriction of a pair of trees to taxon subsets of size $k$,
is it possible to reconstruct the pair of trees?'' This gives a
theorem which extends any identifiability result for a specific pair of trees
of size six to arbitrary pairs of trees under a molecular clock
(Theorem~\ref{thm:recover_clock}). A different approach
shows identifiability of rates-across-sites models for pairs
of trees (Theorem~\ref{nomixthm}). Finally, we show that if a
two class phylogenetic mixture on a single tree mimics the expected site-pattern
frequency vector of a tree on another topology then the two topologies can
differ by at most one nearest neighbor interchange.

\section{Geometry of unbounded mixtures on one or more topologies}

In this section we show that the space of phylogenetic mixtures under the random
cluster model is the convex hull of a finite set of points, i.e. a
convex polytope. The description of the
vertices of the polytope has some interesting consequences discussed
in Section~\ref{sec:polytope}. We then compute dimensions, which is
motivated in part by the following theorem of Carath\'eodory:
\begin{thm}
If $X$ is a $d$--dimensional linear space over the real numbers,
and $A$ is a subset of $X$, then every point of the convex hull of $A$
can be expressed as a convex combination of not more than $d+1$ points
of $A$.
\label{thm:caratheodory}
\end{thm}
A proof can be found as statement 2.3.5 of \cite{grunbaum}. Therefore if we know
that the dimension of a certain set of phylogenetic mixture distributions is $d$,
then any mixture distribution in that set can be expressed as a
phylogenetic mixture with no more than $d+1$ classes.

We also show that the dimension of those site-pattern frequency vectors which can
be written as phylogenetic mixtures on the star tree is equal to
the corresponding dimension for all topologies together. This forms an
interesting contrast to the genericity results in \cite{allman-rhodes}.

Convex polytopes are typically specified in one of two ways:
by a \emph{V-description}, as the convex hull of a finite set of points, or
by an \emph{H-description}, as the bounded intersection of finitely
many half-spaces. Classical algorithms exist to go between the two
descriptions; these are implemented in the software \polymake\
\cite{polymake}. We will make use of both descriptions; for example,
the intersection of polytopes can be easily computed by taking the
union of the two sets of inequalities describing the half-spaces of
the $H$-descriptions.
More introductory material about polytopes can be found in
the texts of Gr\"unbaum \cite{grunbaum} and Ziegler \cite{ziegler}.

From the phylogenetic perspective, we are interested in the set of site
pattern frequency vectors which correspond to non-identifiable
mixtures. In particular, one might ask the question: which
site-pattern frequency vectors can be expressed as a phylogenetic mixture 
on any one of a collection of tree topologies?  At
least in the case of the random cluster model, the answer is the
intersection of the corresponding phylogenetic mixture polytopes.  Using \polymake\
and Proposition~\ref{prop:ST} this becomes an easy exercise for small
trees: simply take the union of the $H$-description inequalities for
the polytope associated with each topology. Although the complexity of
going from a $V$-description to an $H$-description is still open
\cite{polytopeProblems03}, in practice no fast algorithm is known and
so our approach may not feasible for large trees. We analyze the
polytopes associated with quartet trees in Section~\ref{sec:polytope}.

\subsection{The random cluster model}

In this section we define the random cluster model, which generalizes the two
state symmetric (CFN) and Jukes-Cantor DNA models \cite{felsenstein}
in two ways: first, it allows an arbitrary number of states, and
second, it allows non-uniform base frequencies.
We will use the common convention that $[\nstates] := \{1,\ldots,\nstates\}$.
Assume $\nstates$ states, and fix a distribution $\pi = (\pi_i: i \in
[\nstates])$ as the stationary
distribution on those states.
It is always assumed that $\pi_i > 0$ for all $i \in [\nstates]$.
We will label the $n$ states $x_1$, \ldots, $x_n$.

First we define a distribution on site patterns based on partitions.
Informally, we sample once from $\pi$ for each set of the partition,
and assign that value to each element of that set.
\begin{defn}
  \label{defn:DS}
  Let $\S = \{S_1, \ldots, S_r\}$ be a partition of $[n]$. 
  We denote by $D_{\S}$ the
  probability distribution of the random vector $(x_1, \ldots, x_n)$
  obtained by sampling $y_i$ independently from $\pi$ for each $i \in [r]$ and assigning
  state $y_i$ to all of the $x_j$ such that $j \in S_i$.
\end{defn}

We make the following simple observations:
\begin{lem}
  \pushQED{\qed}
Assume $(x_1, \ldots x_n)$ is distributed according to $D_{\S}$. Then
\begin{itemize}
\item
The marginal distribution of each $x_i$ is given by $\pi$.
\item
For all $S \in \S$ and $i,j \in S$ it holds that $x_i = x_j$.
\item
  The collections of random variables $\{x_S : S \in \S\}$ are
  mutually independent, where $$x_S = \{x_i: i \in S\}.$$
\end{itemize}
\popQED
\end{lem}

\begin{defn}
  For any tree $T=(V,E)$ and function $c : E \to [0,1]$,
  define the \emph{random cluster} model as follows:
For each edge $e$ declare the edge 
``closed'' with probability $c(e)$ and declare it ``open'' otherwise.
Let $S_1,\ldots,S_r$ denote the maximal open-edge connected components
of $V$.
Now define the partition $\S = S_1, \ldots, S_r$ and sample a site
pattern from the distribution $D_{\S}$ as in Definition~\ref{defn:DS}.
We use $D_{T,c}$ denote the induced distribution of state assignments to the leaves.
\end{defn}
We will also consider the case $\nstates=\infty$ in which different clusters
will always be assigned different states. Note that this particular case is what
was referred to as the ``random cluster model'' in \cite{mosselSteel04}.

The CFN and Jukes-Cantor DNA models are random
cluster models with $\pi$ the uniform distribution on $2$ and $4$ states
respectively. In general, for any $k$-state model with uniform stationary
frequencies, the corresponding probability in the random cluster model that an edge is closed is $k/(k-1)$ times
the probability of mutation along that edge (see, e.g., \cite{semple-steel} p.197).

\begin{defn}
  A \emph{binary edge vector} is a mapping $g: E \rightarrow \{0,1\}$ taking
  the value 1 if the edge is closed and 0 if the edge is open. 
  % Let $G(E)$ denote the set of all binary edge vectors for a set of edges $E$.
\end{defn}

\begin{defn}
Given an edge probability vector $c: E \rightarrow [0,1]$ let $\jodist{c}$ be the
associated distribution on binary edge vectors, i.e.
\[
J_c(g) = \prod_{e \in E} c(e)^{g(e)} \left(1-c(e) \right)^{1-g(e)}.
\]
\end{defn}

The following lemma can be checked by substituting in the previous definition.
\begin{lem}
  \label{lem:jodist}
  \pushQED{\qed}
If $c_1$ and $c_2$ differ on at most one edge $e$
and if $c = \alpha c_1 + (1-\alpha)c_2$, then
\[
\jodist{c} (g) = \alpha \jodist{c_1}(g) + (1-\alpha) \jodist{c_2} (g).
\]
\popQED
\end{lem}

Let $\vect{x}:[n] \rightarrow [\nstates]$ be an assignment of states to taxa,
i.e. a site pattern.
We can write out the probability of seeing this site pattern under the
random cluster model as
\begin{equation}
  \label{eq:joConditional}
  D_{T,c}(\vect{x}) = \sum_{g} P(\vect{x} | g) \jodist{c} (g)
\end{equation}
where $P(\vect{x} | g)$ is the probability of seeing $\vect{x}$
assuming a binary edge vector $g$. Using this we have

\begin{prop} \label{prop:T01}
For any tree $T$ and any $c$, the distribution $D_{T,c}$ is a convex
combination of distributions $D_{T,c_i}$ where $c_i$ obtains only the values
$0$ or $1$.
\end{prop}
\begin{proof}
  Using Lemma~\ref{lem:jodist} and (\ref{eq:joConditional}) we can proceed
  stepwise: first we obtain (by averaging $D_{T,c_i}$) the set of vectors with the correct first
  coordinate of $c$ and arbitrary other coordinates chosen from
  $\{0,1\}$. Averaging these vectors one can obtain a set of vectors with
  the first two coordinates correct, and so on.
\end{proof}

By grouping all of the open-edge-connected subsets into a partition or
by opening and closing edges according to a partition,
one has the following lemma.
\begin{prop} \label{prop:ST}
  \pushQED{\qed}
Let $T$ be a phylogenetic tree and let $c$ be 
edge probabilities, all of
whose values are in $\{0,1\}$. Then $D_{T,c} = D_{\S}$ for some
partition $\S$ of $[n]$.
On the other hand, for every partition $\S$ of $[n]$ there exists a
phylogenetic tree $T$
and edge probabilities $c \in \{0,1\}^E$ such that
$D_{T,c} = D_{\S}$.
\popQED
\end{prop}

In fact, the distributions $D_{\S}$ determine the convex geometry of
phylogenetic mixtures. 
\begin{thm} \label{thm:convex}
The set of phylogenetic mixtures on trees over $n$ leaves is a
convex polytope with vertices
\[
\{ D_{\S} : \S \mbox{ a partition of } [n] \}.
\]
\end{thm}
{\em Proof.}
The set of phylogenetic mixtures is convex by definition.
By Propositions \ref{prop:T01} and \ref{prop:ST} it follows that every
phylogenetic mixture can be written as a convex sum of the elements
$D_{\S}$. It thus remains to show that we cannot write $D_{\S}$ as a
convex combination of $D_{\S_1},\ldots,D_{\S_k}$ if
$\S \notin \{\S_1,\ldots,\S_k\}$.

Assume by contradiction that
\begin{equation} \label{eq:convex_ext}
D_{\S} = \sum_i \alpha_i D_{\S_i},
\end{equation}
where $\alpha_i > 0$ for all $i$ and $\sum_i \alpha_i = 1$.
\begin{claim}
$\S$ is a refinement of $\S_i$ for all $i$.
\end{claim}
\begin{proof}
Suppose $\S$ does not refine $\S_1$.
Thus there exist $i \neq j$ such that $i$ and $j$ belong to the same
set in $\S$ but do not belong to the same set in $\S_1$.
But this implies by definition that for $D_{\S}$ we have that
$x_i = x_j$ with probability one while for $D_{\S_1}$ the variables
$x_i$ and $x_j$ are independent. This is a contradiction.
\end{proof}

We use $D[f]$ to denote the expectation of
$f$ under the distribution $D$.
The following claim concludes the proof of the theorem.
\begin{claim}
$D_{\S}$ cannot be written as a convex combination of the $D_{\S_i}$.
\end{claim}
\begin{proof}
By the previous claim, we may assume (\ref{eq:convex_ext})
where $\S$ is now a refinement of each of the $\S_i$.
Let $$f(x_1,\ldots,x_n) = \sum_{i,j} 1(x_i = x_j).$$
Note that for a general partition $\S'$ it holds that
\[
D_{\S'}[f] = |\S'|_2^2 + (n^2 - |\S'|_2^2) |\pi|_2^2
\]
where $\left|\S'\right|_2^2 = \sum_{S \in \S'} |S|^2$ and
$\left|\pi\right|_2^2 = \sum_{x \in [\nstates]} \pi_x^2$. In particular, it
follows that since $\S$ is a refinement of $\S_i$ and $\S \neq \S_i$
for all $i$, we have $D_{\S_i}[f] > D_{\S}[f]$ for all $i$. Plugging
this into~(\ref{eq:convex_ext}) we obtain a contradiction. The proof
of the claim follows, thereby completing the proof of
Theorem~\ref{thm:convex}.
\end{proof}

Now we calculate dimensions. The dimension of a convex polytope is
defined to be the dimension of its affine hull.
We do not give a general dimension formula here -- instead we
will just discuss the two state and infinite state models.
We let $\D_n(1/2,1/2)$ denote the space of all distributions that can be
written as a convex combination of phylogenetic trees on $n$
leaves under the CFN model, and let $\Dstar_n(1/2,1/2)$ denote those which
can be written using sets of edge lengths on the star tree with $n$
leaves.

\begin{prop}
\[
\dim(\Dstar_n(1/2,1/2)) = \dim(\D_n(1/2,1/2)) = 2^{n-1} - 1.
\]
\end{prop}
\begin{proof}
We will work with the two-state Fourier transform $F$ as follows.
Because in this case the stationary distribution is uniform, we can
work with ``collapsed'' site-pattern frequency vectors; 
we index these by subsets $B
\subseteq [n-1]$ (see, e.g., \cite{semple-steel}). Now, rather than having the two
states be 0 and 1, take them to be $-1$ and $1$. Thus, the $B$-coordinate
of a site-pattern frequency vector is the probability of having $B$ be
exactly the set of indices $i$ such that $x_i = -1$. Define for any $A
\subseteq [n-1]$ and $D$ any distribution on (collapsed) site-pattern
frequencies
\[
F_A(D) = D\left[\prod_{i \in A} x_i\right].
\]
To see the connection with the Fourier transform defined by a
Hadamard matrix, pick some $B \subseteq [n-1]$ and 
take $D'$ to be the distribution that assigns $-1$ exactly to the
$x_i$ with $i \in B$ (with probability one). Then
\[
F_A (D') = \left( -1 \right)^{|A \cap B|}.
\]
This connection demonstrates that $F$ is invertible. Now, since the Fourier transform is linear and
invertible, and we can compute the dimension of the $\D$'s by computing
the dimension of their image under the Fourier transform.

By definition we have
\begin{equation}
  F_{\emptyset}[D_{T,c}] = 1,
  \label{eq:Femptyis0}
\end{equation}
and it is known that
\begin{equation}
  F_A[D_{T,c}] = 0 \hbox{ for all } A \hbox{ of odd size}
  \label{eq:FSis0}
\end{equation}
for all $T$ and $c$. This last fact can be seen as follows. By
Proposition~\ref{prop:T01} we can assume that $D_{T,c}$ is given by independent
assignment of states (according to $\pi$) to clusters $S_1, \ldots,
S_r$. Because the cardinality of $A$ is odd, at least one of the $A \cap S_j$ must have
odd size, and
\[
D\left[\prod_{i \in A \cap S_j} x_i\right] = -1 \cdot \frac{1}{2} + 1
\cdot \frac{1}{2} = 0.
\]
Equation (\ref{eq:FSis0}) now follows because the expectation of a
product of independent random variables is the product of the
expectations.

It thus follows that equalities (\ref{eq:Femptyis0}) and
(\ref{eq:FSis0})
hold for all distributions in $\D$. This implies that
\[
\dim(\D_n(1/2,1/2)) \leq 2^{n-1} - 1.
\]
We show next that
\begin{equation}
2^{n-1} - 1 \leq \dim(\Dstar_n(1/2,1/2)) \leq \dim(\D_n(1/2,1/2))
  \label{eq:lower}
\end{equation}
which will imply the proposition. The second inequality follows by
containment.

Now we show the first inequality.
Given a set $S$, consider the partition $\rho(S)$ that has the sets $S$ and
a singleton set corresponding to each element of $[n] \setminus S$.
This partition can be achieved on the star tree by declaring all of the
edges in $S$ to be closed with probability one and all of the other
edges to be open with probability one. By the same argument as for 
(\ref{eq:FSis0}),
\[
F_{A}[D_{\rho(S)}] = 1 \mbox{ iff } A \subseteq S \mbox{ and } A \mbox{
  is even}.
\]
Thus $F_{A}[D_{\rho(\emptyset)}]$ is zero for all $A \neq
\emptyset$. It follows
(using the fact that $F_{\emptyset}[D_{T,c}] = 1$ for any $T,c$)
that in this case affine dimension coincides
with linear dimension. Therefore to show the first inequality of
(\ref{eq:lower}) it suffices to find for every
set $S$ of even order a linear combination of elements of $\Dstar_n(1/2,1/2)$ whose Fourier
coefficient at $S$ is $1$ and is $0$ at all other sets.
An inductive argument shows that in order to achieve this task, it
suffices to show that for every even set $S$ there exists an element
of $\D$ whose Fourier coefficient at every even subset of $S$ is
$1$ and is zero on all other sets. This is exactly
$D_{\rho(S)}$ as described above.
The proof follows.
\end{proof}

We now analyze the random cluster for $\nstates=2$ when the distribution
$\pi$ is not uniform. Define $\Dstar_n(r,1-r)$ and $\D_n(r,1-r)$ for
the case of non-uniform $\pi = (r,1-r)$ analogous to the symmetric
(CFN) case for any $0<r<1$.
\begin{prop}
Let $0 < r < 1$ and $r \neq 1/2$. Then
\[
\dim(\Dstar_n(r,1-r)) = \dim(\D_n(r,1-r)) = 2^{n} - n - 1.
\]
\end{prop}
\begin{proof}
  Here we need a variant of the above-described Fourier transform---
  now we take the state space to be $\{r-1,r\}$, with $\pi$ giving the
  first state with probability $r$ and the second state with
  probability $1-r$. Again $F$ will denote the Fourier transform so that
\[
F_A(D) = D\left[\prod_{i \in A} x_i\right].
\]
However, there is one subtle difference, which is that because the
stationary distributions are not uniform, we cannot collapse the
site-pattern frequency vectors. Thus the above $A$ is a subset
of $[n]$, and the coordinates of $D$ are now indexed by subsets of $[n]$. 
The matrix representation of this transform in the $n=1$ case in the
basis $\{\emptyset, \{1\}\}$ is thus
\[
X = 
\left(
\begin{matrix}
  1 & 1 \\
  r & r-1
\end{matrix}
\right).
\]
For $n>1$, the matrix representation is the $n$-fold Kronecker product
of $X$; it follows that this transform is invertible for all $0<r<1$. As before we calculate
the dimension of the Fourier transform of the $\D$.
By definition
\[
F_{\emptyset}[D_{T,c}] = 1,
\]
and if $A$ is a singleton then
\[
F_A\left[D_{T,c}\right] = 0,
\]
for all $T$ and $c$ by a similar argument to before. It thus follows that the equalities above
hold for all distributions in the $\D$. This implies that
\[
\dim(\D_n(r,1-r)) \leq 2^{n} - n - 1.
\]

As before, given a set $S$, consider the partition $\rho(S)$
that has the sets $S$ and
a singleton set corresponding to each element of $[n] \setminus S$.
Then
\[
F_{S}[D_{\rho(S)}] = r \, (r-1)^{|S|} + (1-r) \, r^{|S|} =
r \, (r-1)\left((r-1)^{|S|-1} + r^{|S|-1}\right) \neq 0,
\]
since $0 < r < 1, r \neq 1/2$ and $|S| > 1$.
On the other hand, if $A$ is not a subset of $S$ then
\[
F_{A}[D_{\rho(S)}] = 0
\]
by an argument as in the previous proof.

As before the affine
dimension coincides with the linear dimension.
To prove the corresponding lower bound it suffices to find for every
set $S$ of size at least two a linear combination of elements of
$\Dstar_n(r,1-r)$
whose Fourier coefficient at $S$ is one and is zero at all other sets.
An inductive argument using $D_{\rho(S)}$ again concludes the proof.
\end{proof}

We have just seen how for the CFN model the affine dimension of the
space of phylogenetic mixtures (which has exponential order in $n$) is much smaller
than the number of extremal points (which is the number of partitions
of $n$).
In contrast, for $\nstates=\infty$, the dimension equals the number of
extremal points. This follows from the following proposition.

\begin{prop} \label{prop:qinfty}
  The distributions $D_{\S}$ where $\S$ runs
  over all partitions of $[n]$ are linearly independent.
\end{prop}

\begin{proof}
Recall that in the $\nstates=\infty$ model, each partition is assigned a
different state. 
Thus there is nothing to prove as the probability space we are working in is
the space of partitions of $[n]$.
\end{proof}

\subsection{The phylogenetic mixture polytope for the CFN model}

\label{sec:polytope}

This section specializes to the case of phylogenetic mixtures under
the CFN model. As mentioned previously, the CFN model is equivalent to
the random cluster model with two states and a uniform stationary
distribution. Rather than probabilities of edges being open and
closed, however, it is described in terms of ``branch lengths.''
For a given branch length $\gamma$ we will call $\theta = \exp(-2 \gamma)$
the ``fidelity'' of an edge, which ranges between zero (infinite
length edge) and one (zero length edge) for non-negative branch lengths.
The closed-edge probability $c$ for that edge is then $1-\theta$ which
is twice the probability of a state change along that edge.
\begin{cor}
  The set of phylogenetic mixtures under the CFN model on a given tree
  is a convex set whose extremal points are given (perhaps with
  repetition) by branch length assignments to that topology taken from
  the set $\{0,\infty\}$.
  \label{cor:cfn_extremal}
\end{cor}
\begin{proof}
  A branch length of zero corresponds to an
  edge being open in the random cluster model with probability one,
  and a branch length of infinity
  corresponds to an edge being closed with probability one. The
  corollary now follows from Proposition~\ref{prop:T01}.
\end{proof}

Before analyzing various associated polytopes,
we fix some notation and remind the reader of some facts.
Denote site patterns on $n$ taxa using subsets $A \subseteq
[n-1]$ in the ``collapsed'' notation as before.
Note that one could equivalently use
even sized subsets of $[n]$ via the $f(A)$ below as in \cite{matsen-steel07}.
We will use $p_A$ to denote the probability of a collapsed site
pattern $A$ and $q_A$
to denote the $A$th component of the Fourier transform 
as in \cite{matsen-steel07,semple-steel}.
We will denote the corresponding vectors by $\pvect$ and $\qvect$.
The Hadamard matrices will be denoted $H$; $H$ is symmetric
and $H H = 2^{n-1} I$ when $H$ is $n$ by $n$.
We will denote inner product of $v$ and $w$ by $\ip{v}{w}$ and will
often use the fact that $\ip{Hv}{w} = \ip{v}{Hw}$. We will take $e_A$
to be the vector with $A$'th component one and other components zero.
We will also use the following lemma, from the the proof of Theorem~8.6.3 of
\cite{semple-steel}.
\begin{lem}
\label{lemsip}
  \pushQED{\qed}
  For any subset $A \subseteq \{1,\ldots,n-1\}$ of even
  order, let
  \[
  f(A) = \left\{
  \begin{split}
    & A \hbox{ if } |A| \hbox{ is even}\\
    & A \cup \{n\} \hbox{ otherwise.}
  \end{split}
  \right.
  \]
  Then
  \begin{equation}
    q_A = \prod_{e \in \cP(T,f(A))} \theta(e)
    \label{eq:path_prod}
  \end{equation}
  where $\cP(T,f(A))$ is the unique set of edges which lie in the set of
  edge-disjoint paths connecting the taxa in $f(A)$ to each other.
  \popQED
\end{lem}

We will abuse notation by taking $\Co(T_1,\ldots,T_n)$ to denote the
convex hull of phylogenetic mixtures on trees $T_1, \ldots, T_n$ of
the same number of leaves.

There are four tree topologies on four taxa: the star tree $\tstar$
and the three resolved trees on four taxa $\ttwo$, $\tthree$, and $\tfour$. Thus,
up to isomorphism, there are six convex polytopes of interest in this
case, with inclusions as
indicated:
\begin{eqnarray}
\Co(\tstar) & \subseteq & \Co(\ttwo) \cap \Co(\tthree) \cap
\Co(\tfour) \label{poly:isect} \\
& \subseteq & \Co(\ttwo) \cap \Co(\tthree) \label{poly:twonthree} \\
& \subseteq & \Co(\ttwo) \label{poly:two} \\
& \subseteq & \Co(\ttwo,\tthree) \label{poly:twouthree} \\
& \subseteq & \Co(\ttwo,\tthree,\tfour). \label{poly:union}
\end{eqnarray}
It will be shown below that the inclusion in (\ref{poly:isect}) is an
equality.

From a phylogenetic perspective, polytope (\ref{poly:isect}) represents
those site-pattern frequency vectors which can be realized as a mixture on any
of the four topologies. Polytope (\ref{poly:twonthree}) contains the
distributions from mixtures on two of the resolved
topologies. Polytopes
(\ref{poly:two}), (\ref{poly:twouthree}), and (\ref{poly:union})
correspond to mixtures on one, two, or three resolved topologies.

Polytopes (\ref{poly:isect}) and (\ref{poly:twonthree}) are of special
interest, as they represent mixtures which are non-identifiable for
phylogenetic reconstruction.
In Observations~\ref{obs:tstar_h_rep} and \ref{obs:isect_is_tstar} we
are able to precisely delineate the set of non-identifiable mixtures;
these generalize the non-identifiable mixture examples of 
\cite{matsen-steel07,stefankovic-vigoda}. The drawback is that the
mixtures found here may use as many as eight sets of branch lengths
(recall
Theorem~\ref{thm:caratheodory}) rather than just two, and that we
may mix trees with extreme branch lengths.

There is one more polytope which we will investigate, which is that
cut out by inequalities known to be satisfied for phylogenetic
mixtures.
We will call this polytope $\sens$.
Specifically, $\sens$ is the polytope cut out by $0 \leq q_A \leq 1$
for any $A$, and the Fourier transform of the inequalities $0 \leq p_A
\leq p_{\emptyset}$ for any $A$ and the equality $\sum_{A} p_A = 1$.
Note that the equality is equivalent to $q_{\eset} = 1$.
The inequality $p_A \geq 0$ is equivalent to $\ip{e_A}{\pvect}
\geq 0$ (where $e_A$ is the unit vector defined just prior to Lemma~\ref{lemsip}), and this is equivalent to
\begin{equation}
  \ip{H e_A}{\qvect} \geq 0.
  \label{eq:HeA}
\end{equation}
The following observation notes further redundancies.
\begin{obs}
 $\ip{H e_A}{\qvect} \geq 0$ and $q_A \geq 0$ for every split $A$
 implies $q_\eset \geq q_A$ for every split $A$. These same hypotheses
 also imply that the corresponding probability distribution on splits
 is ``conservative,'' i.e. that $p_\eset \geq p_A$ for any $A$.
\end{obs}
\begin{proof}
  Assume there are $n$ taxa. For the first assertion, let $J$ be the
  $n$ by $n$ matrix with all entries one. Then $J-H$ is a matrix with
  non-negative entries.  Therefore $\ip{H e_A}{\qvect} \geq 0$ for
  every split $A$ implies that $\ip{H (J-H) e_A}{\qvect} \geq 0$ for
  every split $A$. But $HJe_A = H\onevect = 2^{n-1} e_{\eset}$ and $HH
  = 2^{n-1} I$, giving the first assertion.
  For the second assertion, note that $H e_\eset - H e_A$ is a vector
  with non-negative entries, since $H e_\eset$ has all entries equal to $+1$ while $H e_A$ has half its entries
  equal to $+1$ and half equal to $-1$. Thus $\ip{H e_\eset - H e_A}{\qvect}$ is
 non-negative given the assumptions. Thus $\ip{e_\eset - e_A}{\pvect}
  \geq 0$, which is equivalent to the second assertion.
\end{proof}
Because of these observations we note that $\sens$ is the
polytope in Fourier transform space  cut out by $q_A \geq 0$ and
(\ref{eq:HeA}) for each $A$, as well as $q_\eset = 1$.

The following is a simple use of \polymake\ to go from a
$V$-representation to an $H$-representation.
\begin{obs}
  \pushQED{\qed}
  $\Co(\tstar)$ is defined by $q_\eset=1$, $q_{123} \geq 0$ and the
  inequalities (\ref{eq:HeA}) and $q_A \geq q_{123}$ for each $A$.
  \popQED
  \label{obs:tstar_h_rep}
\end{obs}
Another \polymake\ calculation demonstrates
\begin{obs}
  \pushQED{\qed}
  The inclusion in (\ref{poly:isect}) is an equality. In phylogenetic
  terms, the site-pattern frequency vectors obtainable
  as a phylogenetic mixture on a tree for each of the three resolved
  quartet topologies are exactly those obtainable as a phylogenetic mixture 
  on the four taxon star tree.
  \popQED
  \label{obs:isect_is_tstar}
\end{obs}

We can now see what trees sit inside the star tree polytope
$\Co(\tstar)$.
\begin{prop}
  The resolved quartet trees whose site-pattern frequency vectors are obtainable
  as a phylogenetic mixtures on the four taxon star tree are exactly those
  such that the internal branch length is shorter than the sum of
  the branch lengths for any pair of non-adjacent edges.
  \label{prop:whichtreesinstar}
\end{prop}
This proposition may come as a surprise for phylogenetics researchers:
even though a given data set may not have any evidence for a
particular split, the data can appear to be exactly that generated on a tree with an
internal edge which is longer than any of the pendant edges. Said
another way, in order for the vector of expected site-pattern frequencies for a
quartet tree to be identifiable, it is necessary that the internal
edge must be longer than the sum of the branch lengths for a single pair of
non-adjacent pendant edges.

\begin{proof}
  Let $\qvect$ denote the Fourier transform of the site-pattern
  frequency vector for the tree in question, which we assume without
  loss of generality to have topology $12|34$. This $\qvect$ can be
  expressed as a phylogenetic mixture on the star tree exactly when it satisfies
  the conditions in Observation~\ref{obs:tstar_h_rep}.
  Because $\qvect$ is the Fourier transform of a site-pattern
  frequency vector generated on a tree, by the above $q_\eset=1$, $q_{123} \geq 0$, and the inequality
  (\ref{eq:HeA}) is thus satisfied for any $A$.
  Now for each $A \subseteq \{1,2,3\}$ we investigate the consequences of the inequality
  $q_A \geq q_{123}$.
  For $A = \{1\}$, the inequality becomes by (\ref{eq:path_prod})
  \[
  \theta_1 \theta_5 \theta_4 \geq \theta_1 \theta_2 \theta_3 \theta_4
  \Leftrightarrow
  \theta_5 \geq \theta_2 \theta_3.
  \]
  Repeating the process for $A = \{2\}, \{1,3\}, \{2,3\}$ and
  simplifying gives
  \[
  \theta_5 \geq \max
  \{ \theta_1 \theta_3, \theta_1 \theta_4, \theta_2 \theta_3, \theta_2
  \theta_4\}.
  \]
  The cases $A = \{1,2\}, \{3\}$ give $1 \geq \theta_3 \theta_4$
  and $1 \geq \theta_1 \theta_2$, which are trivially satisfied, as is
  the case of $A = \{1,2,3\}$.
  Taking logarithms and dividing by $-2$ gives 
  \[
  \gamma_5 \leq \min
  \{ \gamma_1 + \gamma_3, \gamma_1 + \gamma_4, \gamma_2 + \gamma_3, \gamma_2
  + \gamma_4\}.
  \]
\end{proof}

In the previous section we showed that the dimension of those
site-pattern frequency vectors which can be realized as a phylogenetic mixture 
on the star tree is equal to the dimension of those
pattern probabilities which can be realized as an arbitrary
phylogenetic mixture. This means that given a sample from any nowhere-zero
probability distribution on arbitrary phylogenetic mixtures there is a non-zero
probability of having the sample be realizable from the set of mixture
distributions on the star tree. However, it does not give any
quantitative information. Quantitative answers for this and related
questions for the uniform distribution on site-pattern frequencies can be calculated by using
\polymake\ to calculate volumes. Results are reported in
Table~\ref{tab:volumes}.

For example, assume we uniformly choose a random probability distribution on
patterns obtained by a phylogenetic mixture on a given tree. Then
there is a probability of approximately $0.57$ ($\approx 0.173 /
0.302$) that it is non-identifiable, i.e. that it can be written as a
phylogenetic mixture on another tree. More work on the
relevant geometry is needed to determine if such mixtures pose
problems in the parameter regimes usually found in phylogenetics.

\begin{table}[h]
  \begin{center}
    \begin{tabular}{c|c|c}
      polytope & relative volume (approx.) & absolute volume\\
      \hline
      \cotstar & 0.143 & 5/1008 \\
      \cotwoisect & 0.173 & 13/2160 \\
      \cottwo & 0.303 & 53/5040 \\
      \cotwounion & 0.566 & 11/560 \\
      \counion & 0.909 & 53/1680 \\
      \sens & 1 & 5/144 \\
    \end{tabular}
  \end{center}
  \caption{Relative volumes of the polytopes described in the text.
  The absolute volume is that computed in Fourier transform (i.e. $q$-)
  space.}
  \label{tab:volumes}
\end{table}

\section{Mixtures of two trees}

In this section we specialize to the case of phylogenetic mixtures on
two trees, but we generalize the set of mutation models considered.

\subsection{Combinatorics}

In this section we establish a new combinatorial property that allows pairs of binary phylogenetic trees to be
reconstructed from their induced subtrees of size at most six (Theorem~\ref{disent}). The statistical significance of this result
is described in Corollary~\ref{mixoftwo} and the next section.  We begin with some definitions.

Let $B(X)$ denote the collection of binary phylogenetic $X$--trees (up to isomorphism)
and let $B(X,k)$ denote the subsets of $B(X)$ of size at most $k$.
For $T \in B(X)$ and $Y \subseteq X$, let $\restr{T}{Y}$ denote the
induced binary phylogenetic
$Y$--tree obtained from $T$ by restricting the leaf set to $Y$.
For $\PP = \{T_1, \ldots, T_j\} \in B(X,k)$ let
$\restr{\PP}{Y}: = \{\restr{T_1}{Y}, \ldots, \restr{T_j}{Y}\} \in B(Y,k)$.
We will often stray from standard set theoretical notation when
writing
restrictions, for example $\restr{T}{\{a,b,c,d\}}$ will be written
$\restr{T}{abcd}$.

 We say that a collection $M$ of subsets of $X$ {\em disentangles} $B(X,k)$ if
 one can reconstruct any $\PP$ from the corresponding collection $\{\restr{\PP}{Y}: Y \in M\}$.  This is equivalent to the condition that
for any pair $\PP, \PP' \in B(X,k)$ we have
$$\PP = \PP' \Leftrightarrow \restr{\PP}{Y} = \restr{\PP'}{Y} \mbox{ for all } Y \in M.$$
If in addition, there is a polynomial time (in $|X|$)
algorithm that reconstructs
$\PP$  from the set $\{\restr{\PP}{Y}: Y \in M\}$ we say that $M$ {\em efficiently disentangles} $B(X,k)$.

\begin{figure}[h]
\begin{center}
  \includegraphics[width=4in]{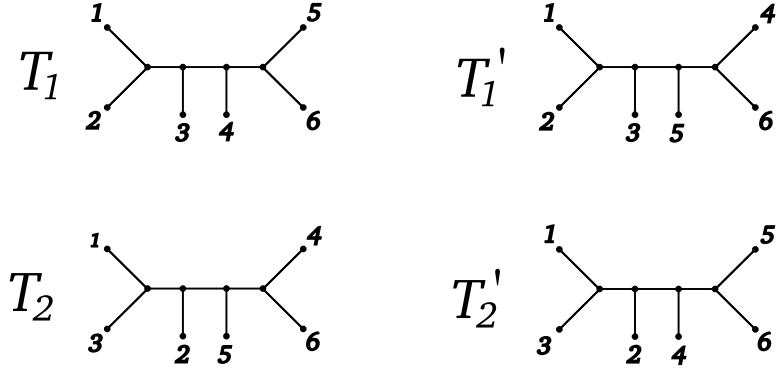}
\end{center}
\caption{Two pairs of trees which have the same combined set of
  splits.}
  \label{hybridfig}
\end{figure}

For example, it is well known that when $k=1$ the collection $M$ of
subsets of $X$ of size four efficiently disentangles $B(X,1)
(=B(X))$; indeed we may further restrict $M$ to just those subsets
of size four that contain a particular element, say $x$, of $X$
(see, e.g., Theorem~6.8.8 of \cite{semple-steel}). However, the subsets of $X$ of size four do not
suffice to to disentangle $B(X,2)$; moreover, neither do the subsets
of $X$ of size at most five. To establish this last claim, let
$X=\{1,2, \ldots, 6\}$, and consider two pairs of trees
shown in Figure~\ref{hybridfig}. Then $\{\restr{T_1}{Y},
\restr{T_2}{Y}\} = \{\restr{T_1'}{Y}, \restr{T_2'}{Y}\}$ for all
subsets $Y$ of size at most five, yet $\{T_1, T_2\} \neq \{T_1',
T_2'\}$.
However, allowing subsets of $X$ of size at most six allows for the
following positive result.

\begin{thm}
\label{disent}
$B(X,2)$ can be efficiently disentangled by the subsets of $X$ of size at most six.
\end{thm}

To establish this result we require the following lemma.

\begin{lem}
\label{thishelps}
Let $T$ be a binary phylogenetic tree on a set $Y$ of seven leaves, and suppose that $S=\{a,b,c\}$ is a subset of $Y$ of size three.
Let $x,y$ be any two distinct elements of $Y - S$.
Then the quartet tree $\restr{T}{S \cup \{x\}}$ is determined by the collection of quartet
trees $\restr{T}{q}$ as $q$ ranges across the following four values:
\begin{itemize}
\item[(i)]
$\{a,b,x,y\}, \{a,c,x,y\}, \{b,c,x,y\}$, and
\item[(ii)]
$\{a,b,c,y\}$.
\end{itemize}
\end{lem}

\begin{proof}
  Consider $\restr{T}{abcy}$. Without loss of generality we may
suppose that $\restr{T}{abcy} = ab|cy$.  If
$\restr{T}{abxy}= ab|xy$ then $\restr{T}{S\cup\{x\}} = ab|cx$. On the other hand,
if $\restr{T}{abxy} = ax|by$ (or $ay|bx$) then $\restr{T}{S\cup\{x\}} = ax|bc$ (or $ac|bx$, respectively).
\end{proof}

\begin{proof}[Proof of Theorem~\ref{disent}]
Consider the collection $Q$ of quartets of $X$ that contain a given element $x \in X$.
The quartets in $Q$ are of two types:  let  $Q_1$ denote the quartets
$q$ in $Q$ for which $\restr{T_1}{q} = \restr{T_2}{q}$ (i.e.
$\restr{\PP}{q}$ consists of just
one tree) and
let $Q_2 = Q-Q_1$. Set $\Q_1:= \{\restr{T_1}{q} \ (= \restr{T_2}{q}): q \in Q_1\}$ and set
$$\Q_2:= \{\restr{T_1}{q}: q \in Q_2\} \cup \{\restr{T_2}{q}: q \in Q_2\}.$$
From $\Q_2$ we construct a graph $G(\Q_2)$ that has vertex set $\Q_2$
and
that has an edge between two quartet
trees, say $ij|kl$ and $i'j'|k'l'$, precisely if one of the trees in
$\PP$ displays both of these quartet trees. Note that
$G(\Q_2)$ is the disjoint union of two cliques.
Moreover, for any two
quartets $q,q' \in Q_2$, each of the two trees in $\Q_2$ that correspond
to $q$ is adjacent (in $G(\Q_2)$) to precisely one of the two trees in
$\Q_2$ that correspond to $q'$, and the resulting two edges form a matching for
these four vertices.

Now, provided $q \cup q'$ has
cardinality at most six we can determine this matching since we can, by
hypothesis,  construct
$\restr{\PP}{q \cup q'}$ which
must consist of two trees, and
this pair of trees tells us how to match the two resolutions  provided
by $\PP$ for $q$  (viz. $\{\restr{T_1}{q}, \restr{T_2}{q}\}$)
with the two resolutions of $q'$ (viz. $\{\restr{T_1}{q'},
\restr{T_2}{q'}\}$).  In particular we can determine the two
edges of $G(\Q_2)$ that connect these four vertices of $G(\Q_2)$.

We claim that we can also  determine (in polynomial time using just
$\restr{\PP}{Y}$ for
choices of $Y$ of size at most $six$)  the matching
between these four vertices of $G(\Q_2)$ in
the remaining case where $q \cup q'$ has cardinality seven.

Accepting for moment this claim, this
 allows us to reconstruct all the edges of $G(\Q_2)$ and in particular the two
disjoint cliques of $G(\Q_2)$, which bipartition $\Q_2$. Taking the
union of each clique with $\Q_1$ provides the pair of subsets
$\{\{\restr{T_1}{q}: q \in Q\}, \{\restr{T_2}{q}: q \in Q\}\}$
from which  $\{T_1, T_2\}$ can be recovered. Furthermore all of this
can be achieved in polynomial time.

Thus it remains to establish the claim.
Take two quartets $q = \{a,b,c,x\}$ and $q'= \{a',b',c',x\}$ from $Q_2$
where we are assuming (since $|q \cup q'|=7$) that
  $$\{a,b,c\} \cap \{a',b',c'\} = \emptyset.$$
We will now invoke
Lemma~\ref{thishelps} with $S = \{a,b,c\}$ and $Y = q \cup q'$.
Assume all of the four quartets in Lemma~\ref{thishelps} are in $Q_1$;
by the conclusion of the lemma the quartet tree $\restr{T}{abcx}$ is
uniquely determined. Thus $\{a,b,c,x\} \in Q_1$, which contradicts our
assumption. Therefore at least one of
the four quartets of type  (i) or (ii) in Lemma~\ref{thishelps} is
in $Q_2$.

Suppose there exists a
a quartet $q^*$ of type (i) in Lemma~\ref{thishelps}.
 Then $q \cup q^*$ and $q' \cup q^*$ both have cardinality at most six
 (for the latter, note that
 $y$ in Lemma~\ref{thishelps} must be one of the elements $a',b',c'$
 as $y \in Y - q$) and so
we can determine the matching. Similarly, since $\{a',b',c',x\} \in Q_2$
we can invoke Lemma~\ref{thishelps} with $S=\{a',b',c'\}$ and the pair
$x,y'$ where $y'$ is an element of $Y-S$ different from $x$. By
similar logic, at least one of the quartets satisfying condition (i) or (ii) in
Lemma~\ref{thishelps} must also be in $Q_2$ for this choice of $S$. Once again if we can find a
quartet satisfying condition (i) of Lemma~\ref{thishelps} we can
determine the matching. A remaining possibility is that in both
cases (i.e. for $S= \{a,b,c\}$  and $S = \{a',b',c'\}$) we can only
find a quartet in each case that satisfies condition (ii) of Lemma~\ref{thishelps}. Call these
two quartets $q_1 =\{a,b,c,y\}$ and $q_1'=\{a',b',c',y'\}$,
respectively.
 Then the three sets
$q \cup q_1$, $q' \cup q_1'$ and $q_1 \cup q_1'$ each have cardinality
at most $6$ (for the last case, note that $y'$ is one of $a,b,c$ and $y$
is an element of $a',b',c'$) and so we
can determine the matching for these three pairs. This allows
construction of $\restr{T_i}{q \cup q' \cup q_1 \cup q_1'}$ for $i =
1,2$ from the corresponding quartet trees;
the matching for the four vertices of $G(\Q_2)$ corresponding
to $q \cup q'$ are then available by restriction. This completes the proof.
\end{proof}

An immediate consequence of Theorem~\ref{disent} is the following.

\begin{cor}
\label{mixoftwo}
Suppose a model has the property that from an arbitrary mixture of
 processes on
two trees with the same leaf set of size six we can reconstruct the topology
of the two trees. Then the same property applies for phylogenetic
mixtures on two
trees for any leaf set $X$ (of any size greater than six),
and by an algorithm that is polynomial in $|X|$.
\end{cor}

\noindent {\bf Remarks} \ Peter Humphries has extended Theorem~\ref{disent} to
obtain analogous results for $B(X,k)$ for $k>2$ (manuscript in preparation.)

The algorithm for disentangling two trees outlined in the
proof of Theorem~\ref{disent} would run in polynomial time, and a
straightforward implementation of the method would have a run time
complexity of $O(|X|^7)$. However, it is quite possible that a more
efficient algorithm could be developed for this  problem (and
thereby for Corollary~\ref{mixoftwo}).

\subsection{Models}
\subsubsection*{Clocklike mixtures}

Suppose one has a phylogenetic mixture on two trees $T_1$ and $T_2$. In
this section we are interested in whether one can reconstruct the
pair $\{T_1, T_2\}$ (or some information about this pair) from
sufficiently long sequences.
In the case where for each tree there is a stationary reversible
Markov process (possibly also with rate variation across sites), and the (positive, finite) branch lengths of $T$
satisfy a molecular clock some positive results are possible.

\begin{obs}
The union of the splits in two trees $T_1$ and $T_2$ on the same taxon
set can be recovered from
a phylogenetic mixture on the two trees under a molecular clock.
\label{obs:gotsplits}
\end{obs}

To see this we simply consider the function $p:X \times X
\rightarrow [0,1]$ defined by setting $p(x,y)$ to be the probability
that species $x$ and $y$ are assigned different states by the
mixture distribution (i.e. $p(x,y)$ is the expected normalized Hamming distance
between the sequences).  Then $p = d_1+d_2$
where (by the molecular clock assumption) $d_1$ and $d_2$ are
monotone transformations of tree metrics realized by $T_1$ and
$T_2$ respectively. By split decomposition theory (\cite{dressband}) it
follows that $\Sigma(T_1) \cup \Sigma(T_2)$ can be recovered from
$p$.

Note that $\Sigma(T_1) \cup \Sigma(T_2)$ does not determine the
set $\{T_1, T_2\}$ as the two pairs of trees in
Figure~\ref{hybridfig} shows.
However this example is somewhat special:
\begin{lem}
\label{splitsunt}
Suppose $\{T_1, T_2\}$ and $\{T_1', T_2'\}$ are two pairs of
binary phylogenetic trees on the same set $X$ of six  leaves, and that
$$\Sigma(T_1) \cup \Sigma(T_2) = \Sigma(T_1') \cup \Sigma(T_2').$$
Then either $\{T_1, T_2\} = \{T_1', T_2'\}$ or the two pairs of
trees are as shown in Figure~\ref{hybridfig} (up to symmetries).
\end{lem}

\begin{proof}
  The proof is simply a case-by-case check of split compatibility
  graphs. A split compatibility graph is a graph where each split is
  represented by a vertex and an edge connects two splits which are
  compatible. In this case there are three nontrivial splits for each
  tree topology; three splits being realizable on a tree is equivalent
  to those three splits forming a clique in the split compatibility
  graph. Thus the lemma is equivalent to saying that up to symmetries
  there is only one subset of the vertices of the split compatibility
  graph for six taxa which can be expressed as
  two three-cliques in two different ways.

  There are two unlabeled topologies on binary trees of six leaves: the
  caterpillar (with symmetry group of size eight) and the symmetric
  tree (with symmetry group of size 48). First we divide the
  problem into the case of two caterpillar topologies, then the case
  of one caterpillar and one symmetric topology, finally two
  symmetric topologies. We label the two types of splits as follows:
  we call a split with three taxa on either side (such as $123|456$)
  ``type $x$'', and a split with two taxa on one side and four on the
  other (such as $12|3456$) ``type $y$.''

  Assume $\{T_1, T_2\} \neq \{T_1', T_2'\}$.
  In the case of two caterpillar topologies it can be seen by
  eliminating cases that $T_1$ and $T_2$ cannot share a split of type $y$.
  Therefore the four type $y$ splits of $T_1$ and $T_2$ must form a
  square of distinct vertices in the split compatibility graph.
  Further elimination shows that the two trees in
  Figure~\ref{hybridfig} are the only ones possible up to symmetries.

  The cases involving a symmetric tree are even easier, as the choice
  of two splits in a symmetric tree determines the third. In the case
  of one caterpillar and one symmetric topology, this implies that
  there can be at most four type $y$ splits in $T_1$ and $T_2$.
  Checking cases quickly eliminates all possibilities. Similar
  reasoning deals with the two symmetric topology case, proving the
  lemma.

\end{proof}

\begin{thm}
Suppose that for a reversible stationary model (possibly with rate
variation across sites) there is a method that is able to distinguish
a phylogenetic mixture on trees $T_1$ and $T_2$ from a
phylogenetic mixture on trees
$T_1'$ and $T_2'$ (see Figure~\ref{hybridfig}) under branch lengths
that satisfy a molecular clock on each tree.  Then from any
phylogenetic mixture on
two binary trees for a leaf set $X$ with both sets of branch lengths
subject to a clock, one can recover the two trees by an algorithm that
runs in polynomial ($O(|X|^7)$) time.
\label{thm:recover_clock}
\end{thm}
\begin{proof}
Combine Theorem~\ref{disent},
Observation~\ref{obs:gotsplits}, and Lemma~\ref{splitsunt}. For the
time efficiency estimate, the distance matrix can be estimated in
$O(|X|^2)$ time, and the split decomposition can be done in $O(|X|^4)$
time \cite{dressband}.
\end{proof}

\subsubsection*{Non-clocklike mixtures}

In \cite{matsen-steel07} it was shown that under two-state symmetric
(CFN) model one can have a mixture
of two processes on one tree giving the same 
site-pattern frequency vector as a single process on a different
tree.
This requires that the two sets of branch lengths being mixed
to be quite different and carefully adjusted. For example, we have:
\begin{cor}
  If a two class phylogenetic mixture on a tree $R$ has the same
  site-pattern frequency vector as a tree of a different
  topology $S$, then the two sets of branch lengths cannot be
  clock-like (even for different rootings of the tree), nor can one
  branch length set be a scalar multiple of the other.
\end{cor}
\begin{proof}
  There must be a taxon set $abcd$ such that $\restr{R}{abcd} = ab|cd$
  and $\restr{S}{abcd} = ac|bd$. Using the notation of
  \cite{matsen-steel07}, (also explained in Section~\ref{sec:mimic}) clocklike mixtures must have a pair of
  adjacent taxa (say $a$ and $b$) such that $k_a = k_b$. For one set
  of branch lengths to be a nontrivial scalar multiple of another, all
  of the pendant $k_i$'s must be either less than or greater than one.
  Either of these cases contradicts Proposition~7 of
  \cite{matsen-steel07}.
\end{proof}

However, one could ask if a more complex
phylogenetic mixture on a tree could mimic an unmixed process
on a different tree. Again a molecular clock rules this out, and for
branch lengths that scale proportionately (as in a rates-across-sites
distributions) we now show that identifiability of the underlying
tree still holds.

\begin{thm}
\label{nomixthm}
Consider two
binary phylogenetic trees $T$ and $T'$ on the same leaf set $X$ of
size $n$ generating data under the CFN model. For $T$ suppose we have a mixture of such processes that can
be described by a set of branch lengths and a distribution $\D$ of
rates across sites which generates the same distribution on site
patterns as that produced by an (unmixed)
set of branch lengths on $T'$.  Then $T=T'$ and $\D$ is the degenerate distribution
that assigns all sites the same rate.
\end{thm}
\begin{proof}
It suffices to prove the result for $n=4$ and $X = \{1,2,3,4\}$,
with
$T$ the tree $12|34$, and $T'$ the tree $13|24$.  We denote the edge of
$T$ (resp. $T'$) that is incident with leaf $i$ by $e_i$
(resp. $e_i'$) and the interior edge of $T$ (resp. $T'$) by $e_0$
(resp. $e_0'$). Let  $\theta_i':= 1-2p(e_i')$ and let $\lambda_i$ denote the
branch length of edge $e_i$ so that the probability of a change along
$e_i$ is $\frac{1}{2}(1-f(2\lambda_i))$ where $f(x) =
\EE_{\D}[\exp(\mu x)]$ is the moment generating function for the
distribution of the rate parameter $\mu$ in $\D$.

Then we have (see, e.g., Lemma~8.6.4 and Theorem~8.8.1 of
\cite{semple-steel}):
\[
f(-2\lambda_1-2\lambda_2-2\lambda_0) = \theta_1'\theta_2' \hbox{ and }
f(-2\lambda_3-2\lambda_4-2\lambda_0) = \theta_3'\theta_4',
\]
and thus
$$f(-2\lambda_1-2\lambda_2-2\lambda_0)\cdot f(-2\lambda_3 - 2\lambda_4-2\lambda_0)
=\theta_1'\theta_2'\theta_3'\theta_4'.$$
Also,
$$\theta_1'\theta_3'\theta_2'\theta_4' = f(-2\lambda_1-2\lambda_2-2\lambda_3-2\lambda_4).$$
Combining these last two equations and setting $r:= -2\lambda_1-2\lambda_2, s:
= -2\lambda_3-2\lambda_4$;
\begin{equation}
\label{ineqq}
f(r+s) = f(r-2\lambda_0)f(s-2\lambda_0) \leq f(r)f(s),
\end{equation}
with equality precisely if $\lambda_0 = 0$.
However, $\exp(\mu x)$ is an increasing function of $\mu$ for positive
$x$. It follows that the random variables $\exp(\mu r)$ and $\exp(\mu
s)$ are positively correlated, i.e.
$$f(r+s) \geq f(r)f(s)$$
with equality precisely if $\D$ is a degenerate distribution.  Consequently,
(\ref{ineqq}) is an equality; thus $\D$ is a degenerate distribution
and $T'= T$.
\end{proof}

\noindent {\bf Remark}  Theorem~\ref{nomixthm} extends to provide an analogous
result for the uniform distribution random cluster model on
any even number $q=2r$ of states, since such a model induces the
random cluster model on two states by partitioning the $2r$ states
into two sets, each of size $r$.

\subsection{Mixed branch repulsion: larger trees}
\label{sec:mimic}

In this section we find results analogous to those in
\cite{matsen-steel07} for trees larger than quartet trees. The main
result is that two class phylogenetic mixtures on a tree can only mimic a
tree which is topologically one nearest neighbor interchange away from the original tree.

Let $\leaves(T)$ denote the set of leaves of a given tree $T$. We will write $R
\mixarr S$ to mean that there exists a two class phylogenetic mixture 
on $R$ which gives exactly the same site-pattern
frequency vector as some branch length set on a tree of topology $S$ under
the CFN model. Of course, if $R \mixarr S$ then $\leaves(R) =
\leaves(S)$.
\begin{thm}
  Assume $R$ and $S$ are two topologically distinct trees on at least four
  leaves such that $R \mixarr S$. Then $R$ and $S$ differ
  topologically by one nearest neighbor interchange (NNI). Furthermore,
  assume the NNI partitions $\leaves(R)$ into the sets
  $X_1,\ldots,X_4$. Then $\restr{R}{X_i} = \restr{S}{X_i}$ for any $i$
  (equality as rooted trees with branch lengths).
  \label{thm:one_nni}
\end{thm}

For this proof we will draw notation and several ideas from the proof of the main
result of \cite{matsen-steel07}. For a four taxon tree with taxon labels $1$ through $4$
we will label the the pendant edges with the corresponding numbers. We will write
the quartet tree with the $ab|cd$ split as simply $ab|cd$. Given two sets
of branch lengths on a given tree we use $k_i$ to denote the ratio of the
fidelities (see Section~\ref{sec:polytope}) of the two branch lengths
for the edge $i$. We will constantly use the simple fact that if the
edge of an induced subtree consists of a sequence of edges then the
induced $k_i$ for that edge consists of the product of the $k_i$'s for
the sequence of the edges (this holds because the fidelities are multiplicative along a path, and therefore their ratios are also).

\begin{lem}
  \pushQED{\qed}
  The quartet splits $ab|cd$, $ac|bd$ and $ad|bc$ are invariant under
  the action of the Klein four group
  \[
  K_4 = \{1, (ab)(cd), (ac)(bd), (ad)(bc)\}.
  \]
  \popQED
  \label{lem:k4_invar}
\end{lem}

The following lemma can be checked by hand.
\begin{lem}
  \pushQED{\qed}
  Given numbers $k_a, k_b, k_c$, there exists $\sigma \in K_4$
  such that
  \[
  k_{\sigma(a)} \geq k_{\sigma(b)} \hbox{ and } k_{\sigma(a)} \geq k_{\sigma(c)}.
  \]
  \popQED
  \label{lem:k4_ineq}
\end{lem}

The following lemma is a rephrasing of Proposition~3 of
\cite{matsen-steel07}:
\begin{lem}
  \pushQED{\qed}
  If $ab|cd \mixarr ab|cd$ then the following two statements must be satisfied:
  \begin{itemize}
    \item $k_a = k_b$ or $k_c = k_d$
    \item $k_a = k_b^{-1}$ or $k_c = k_d^{-1}$.
  \end{itemize}
  \popQED
  \label{lem:same_top}
\end{lem}

\begin{lem}
  If $ab|cd \mixarr ac|bd$ then
  \begin{itemize}
    \item There is some element $\sigma \in K_4$ such that
      $k_{\sigma(a)} > k_{\sigma(c)} > k_{\sigma(d)} > k_{\sigma(b)}$
    \item none of $k_a, \ldots, k_d$ are equal to one
    \item either exactly one or exactly three of $k_a, \ldots, k_d$ are greater
      than one
    \item $k_a \neq k_b^{-1}$ and $k_c \neq k_d^{-1}$.
  \end{itemize}
\label{lem:ineq_chain}
\end{lem}

\begin{proof}
  Each item in the list is from Proposition~7 of \cite{matsen-steel07}
  with the exception of the last one. By Lemma~\ref{lem:k4_invar} we
  can relabel such that $k_a > k_c > k_d > k_b$.
  Let $f(x) = \frac{x^2 - 1}{x}$. Note that $f(x^{-1}) = -f(x)$,
  $f(x)$ is positive for $x \geq 1$ and strictly increasing for $x >
  0$. By equation (12) of \cite{matsen-steel07},
  \[
  f(k_a) f(k_d) + f(k_b) f(k_c) > 0.
  \]
  Assume first that $k_a > k_c > k_d > 1 > k_b$. Then the
  above properties of $f$ imply the following deductive chain:
  \begin{eqnarray*}
    f(k_a) f(k_c) + f(k_b) f(k_c) & > & 0 \\
    f(k_a) + f(k_b) & > & 0 \\
    f(k_a) & > & f(k_b^{-1}),
  \end{eqnarray*}
  implying $k_a \neq k_b^{-1}$. The case where
$k_a > 1 > k_c > k_d > k_b$ is similar, as is the proof that $k_c \neq
k_d^{-1}$.
\end{proof}

The proof of Theorem \ref{thm:one_nni} rests on the following 
observation.
\begin{lem}
  \pushQED{\qed}
  If $R \mixarr S$ then $\restr{R}{F} \mixarr \restr{S}{F}$ for any $F
  \subset \leaves(R)$.
  \popQED
  \label{lem:restriction}
\end{lem}

\begin{figure}
  \begin{center}
    \includegraphics[width=2in]{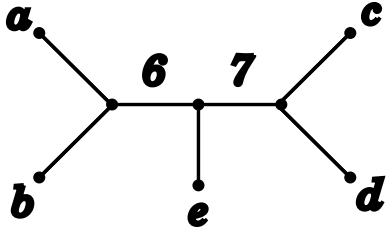}
  \end{center}
  \caption{Definition of $W_{abcde}$.}
  \label{fig:fiveleaf}
\end{figure}

We will use this lemma by restricting taxon sets of the larger tree to
sets of size five, then analyzing for which ordered pairs $(R,S)$ of five leaf
subtrees it holds that $R \mixarr S$. There are 225 ordered pairs of five leaf trees,
however in the following lemma we show that symmetry considerations
reduce the relevant number of interest to four.
For ease of notation, we will write the five leaf subtree $W_{abcde}$
as shown in Figure~\ref{fig:fiveleaf}.

\begin{lem}
  Given trees on five leaves $R$ and $S$, the question of whether $R
  \mixarr S$ or not is equivalent to the question of if one of the
  following is true:
  \begin{eqnarray}
    W_{12345} & \mixarr & W_{12345} \label{mix:zero} \\
    W_{12345} & \mixarr & W_{13245} \label{mix:one} \\
    W_{12345} & \mixarr & W_{12354} \label{mix:two} \\
    W_{12345} & \mixarr & W_{13254} \label{mix:three}
  \end{eqnarray}
\end{lem}
\begin{proof}
  It can be assumed that $R$ is $W_{12345}$ by renumbering.
  Note that the symmetries of a five leaf tree are generated by
  $(12)$, $(34)$, and $(13)(24)$ on the tree $W_{12345}$. A
  combination of these symmetries applied to $R$ and renumbering means
  that these symmetries can then be applied to the labels of $S$ while still
  assuming that $R$ is $W_{12345}$. Using these symmetries
  $S$ can be assumed to be either $W_{abcd4}$ or $W_{abcd5}$. There
  are six such trees; a further application of the symmetries shows
  that the cases of $S = W_{13254}$ and $S = W_{23154}$
  are equivalent, as are $S = W_{13245}$ and $S = W_{14235}$.
\end{proof}

\begin{lem}
  Mixture (\ref{mix:one}) is impossible, i.e. $W_{12345} \nomixarr
  W_{13245}.$
  \label{lem:no_cut_nni}
\end{lem}

\begin{proof}
  Assume the contrary, and that $k_i$'s are labeled as in
  Figure~\ref{fig:fiveleaf}. By (clear extensions of) Lemmas \ref{lem:k4_invar} and
  \ref{lem:k4_ineq} we can assume that $k_1 \geq k_2$ and $k_1 \geq
  k_3$ on these trees. By restricting to the taxon set to 1234, and
  noting that by Lemma~\ref{lem:restriction} $12|34 \mixarr 13|24$, we
  have $k_1 > k_3 > k_4 > k_2$ and that $k_3$ and $k_4$ are
  either both greater than one or both less than one by
  Lemma~\ref{lem:ineq_chain}.  By restricting to $1235$, it is clear
  that $k_5 \neq 1$. Assume $k_5 < 1$. Restricting the taxon set to
  2345 means that $25|34 \mixarr 24|35$; by testing elements of $K_4$
  in Lemma~\ref{lem:ineq_chain} and using the fact that $k_3$ and
  $k_4$ are either both greater than one or both less than one and
  that $k_5 < 1$, one must have $k_2 k_6 > k_4 > k_3 > k_5$. This
  contradicts the above statement that $k_3 > k_4$. The case where
  $k_5 > 1$ follows similarly by restricting to $1345$.
\end{proof}

\begin{lem}
  Mixture (\ref{mix:three}) is impossible, i.e. $W_{12345} \nomixarr W_{13254}.$
  \label{lem:no_sub_q}
\end{lem}

\begin{proof}
  Assume the contrary.
  First restrict to the taxon set $1345$. For this taxon set $15|34
  \mixarr 13|45$, showing by Lemma~\ref{lem:ineq_chain} that $k_3 \neq k_4$, $k_3 \neq k_4^{-1}$,
  and $k_5 \neq 1$. Second, restrict to taxon set $2345$.
  For this taxon set the induced mixture is $25|34 \mixarr 25|34$,
  therefore we can apply
  Lemma~\ref{lem:same_top}. Because $k_3 \neq k_4$ and $k_3 \neq
  k_4^{-1}$, it must be true that $k_2 k_6 = k_5$ and $k_2 k_6 =
  k_5^{-1}$. This contradicts the fact that $k_5 \neq 1$.
\end{proof}

Therefore we are left with mixtures (\ref{mix:zero}) and
(\ref{mix:two}), implying the following corollary.

\begin{cor}
  \pushQED{\qed}
  Assume $R \mixarr S$ for two five-leaf trees $R$ and $S$. Then $R$
  and $S$ share a nontrivial split.
  \label{cor:must_be_nni}
  \popQED
\end{cor}

We now present two more lemmas which will be used in the proof of
Theorem~\ref{thm:one_nni}.  Given rooted trees $R$ and $S$ let
$\ujoin{R}{S}$ denote the unrooted tree obtained by joining the roots
of $R$ and $S$ together with an edge.

\begin{lem}
Assume $\ujoin{R_1}{R_2} \mixarr \ujoin{S_1}{S_2}$, $\leaves(R_1) =
\leaves(S_1)$, and all of the $k$'s for
the edges in $R_1$ are one.  Then $R_1 = S_1$ (equality with
branch lengths).
  \label{lem:ks_one_same}
\end{lem}

\begin{proof}
  Add a taxon $e$ at the root of $R_1$ (resp. $S_1$) to obtain the unrooted
  tree $R_U$ (resp. $S_U$). We will show that the between-leaf distance
  matrices for $R_U$ and $S_U$ are the same, which implies that $R_U = S_U$
  and thus $R_1 = S_1$. Pick $c$ and $d$ distinct in $\leaves(R_2)$.
  Pick an arbitrary $a$ and $b \in \leaves(R_1)$ and
  restrict to the taxon set $abcd$.  By Proposition~4 of
  \cite{matsen-steel07}, the pairwise distance between $a$ and $b$ in $R_1$ and $S_1$ (and
  thus in $R_U$ and $S_U$) will be the same. To show that distances from
  taxa $a \in \leaves(R_1)$ to the root taxon $e$ are the same in $R_U$
  and $S_U$, repeat the same process but for any $a$ choose $b$ such
  that the MRCA of $a$ and $b$ in $R_1$ is the root of $R_1$. Another
  application of Proposition~4 of \cite{matsen-steel07} in this case
  proves the proposition.
\end{proof}

\begin{lem}
  If $\ujoin{R_1}{R_2} \mixarr \ujoin{S_1}{S_2}$, $\leaves(R_1) =
  \leaves(S_1)$ and $\splits(R_2) \neq \splits(S_2)$
  then $R_1 = S_1$ (equality with branch lengths.)
  \label{lem:non_nni_eq}
\end{lem}

\begin{proof}
  For $x,y \in \leaves(R_2)$, let $C_y (x)$ be the set of edges in the
  path from $x$ to the MRCA of $x$ and $y$. Define
  \[
  \varphi_y (x) = \prod_{e \in C_y (x)} k_e.
  \]
  This takes the place (for induced subtrees) of a single $k_e$.
  The idea of the proof is to use the previous
  lemma by showing that $k_e$ for any edge $e$ in $R_1$ is one. However,
  by induction it is enough to show that $\varphi_y (x) = \varphi_x (y) = 1$
  for any $x,y \in \leaves(R_2)$.

  Since $\splits(R_2) \neq \splits(S_2)$ but
  $\leaves(R_2) = \leaves(S_2)$ there exists a subset $\{a,b,c\} \subset
  \leaves(R_2)$ such that $R_2$ restricted to the taxon set $abc$ is the
  tree $(ab)c$, while $S_2$ restricted to $abc$ is $(ac)b$. Pick any
  $x, y \in \leaves(R_1)$. First restrict to taxon set $abcx$, for which
  $ab|cx \mixarr ac|bx$. By Lemma~\ref{lem:ineq_chain}, $\varphi_b (a)
  \neq \varphi_a (b)$ and $\varphi_b (a) \neq [\varphi_a (b)]^{-1}.$
  Now restrict to the taxon set $abxy$, for which $ab|xy \mixarr
  ab|xy$. By Lemma~\ref{lem:same_top}, $\varphi_y (x) = \varphi_x (y)$ and
  $\varphi_y (x) = [\varphi_x (y)]^{-1}$, implying that each $\varphi$ is one.
  The lemma now follows.
\end{proof}

The final lemma allows for the combination of splits; it is a special case of
Lemma~2 of \cite{mea83}. We present an argument here for completeness.
\begin{lem}
  Let $T$ be a phylogenetic tree.
  If $A\cup\{x\} | B \in \splits(\restr{T}{A \cup B \cup \{ x \}})$
  and $A\cup\{y\} | B \in \splits(\restr{T}{A \cup B \cup \{ y \}})$
  then
  $A\cup\{x,y\} | B \in \splits(\restr{T}{A \cup B \cup \{x,y\}})$.
  \label{lem:splitcombine}
\end{lem}
\begin{proof}
  First we note that if $A | B \in \splits(\restr{T}{A \cup B})$ then
  one of $A|B\cup\{x\}$ or $A\cup\{x\} | B$ is contained in
  $\splits(\restr{T}{A \cup B \cup \{ x \}})$, otherwise
  the restriction of $\restr{T}{A \cup B \cup \{ x \}}$
  to $A \cup B$ cannot contain the split $A | B$.

  Applying this fact to the two splits $A\cup\{x\} | B$ and
  $A\cup\{y\} | B$ implies either the conclusion of the lemma or that
  $A\cup\{x\} | B\cup\{y\}$ and $A\cup\{y\} | B\cup\{x\}$ are both in
  $\splits(\restr{T}{A \cup B \cup \{x,y\}})$. This latter option is
  excluded by split compatibility.
\end{proof}

\begin{figure}
  \begin{center}
    \includegraphics[width=2.5in]{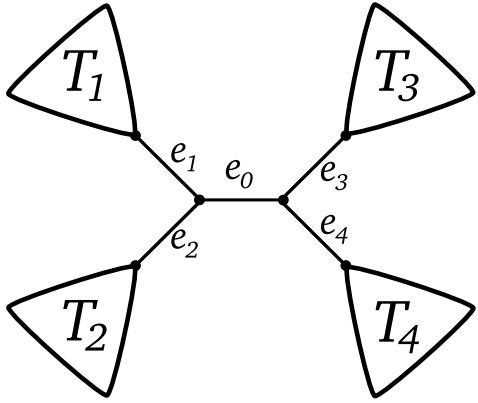}
  \end{center}
  \caption{Notation used in the proof of Theorem~\ref{thm:one_nni}.}
  \label{fig:bigtree}
\end{figure}

\begin{proof}[Proof of Theorem~\ref{thm:one_nni}]
  Because $R$ and $S$ are topologically distinct yet have the same
  number of leaves, there must be at least one split in $R$ which is not in $S$.
  Say this split is given by the edge $e_0$. The edge $e_0$ must
  induce a nontrivial split, and therefore assign $e_1, \ldots, e_4$
  and $T_1, \ldots, T_4$ such that $R$ can be drawn as in
  Figure~\ref{fig:bigtree}.

  Pick any $i \in \{1,\ldots,4\}$. We claim that the split induced by
  edge $e_i$ is in $\splits(S)$. If $|\leaves(T_i)| = 1$ then there is
  nothing to prove, so assume that $|\leaves(T_i)| \geq 2$. Construct
  a five-leaf tree by choosing two leaves $a,b$ from $\leaves(T_i)$ and also
  leaves $c,d,e$: one from each of the other three $T_j$. Because the split
  induced by $e_0$ is not in $S$ by hypothesis, it also cannot be in
  $\restr{S}{abcde}$.  An application of Corollary~\ref{cor:must_be_nni} now
  implies that the split induced by $e_i$ must be in
  $\splits(\restr{S}{abcde})$.
  This is true for each such choice of $abcde$:
  of these choices combined via Lemma~\ref{lem:splitcombine} show that the split induced by the edge
  $e_i$ is in $\splits(S)$.

  Four applications of Lemma~\ref{lem:non_nni_eq} now prove the theorem.
\end{proof}

The following proposition says that the sort of mixture described in
Theorem~\ref{thm:one_nni} is possible (assuming the main result of
\cite{matsen-steel07}). It is a simple general fact.
\begin{prop}
  Let $T_1, \ldots, T_4$ be rooted trees and $R$ and $S$ two trees on
  the taxon set $\{1,2,3,4\}$. Let $\tilde{R}$ and $\tilde{S}$ be the trees
  obtained from $R$ and $S$ by attaching tree $T_i$ in place of taxon $i$.
  Now if $R \mixarr S$ then $\tilde{R} \mixarr \tilde{S}$.
\end{prop}

\begin{proof}
  Let the vector $\vect{y}$ represent the state vector for the
  terminal taxa on $R$ and $S$ and let $\vect{x_i}$ represent the
  state vector for the tree $T_i$.
  Let $p_{\gamma}^T(\vect{z})$ mean the probability of state vector
  $\vect{z}$ on a
  tree $T$ with branch lengths $\vect{\gamma}$; $\vect{\gamma}$ will be omitted if
  understood.
  The statement $R \mixarr S$ means exactly that there exist
  $\vect{\gamma_1}$, $\vect{\gamma_2}$, $\vect{\gamma_3}$ and $\alpha$ such that
  \[
  \alpha p_{\vect{\gamma_1}}^R(\vect{y}) + (1-\alpha)
  p_{\vect{\gamma_2}}^R(\vect{y}) = p_{\vect{\gamma_3}}^S(\vect{y})
  \]
  for any state vector $\vect{y}$. We observe that
  \[
  p^{\tilde{W}}(\vect{x_1},\ldots,\vect{x_4}) =
  \sum_{\vect{y}} \, p^W(\vect{y}) \prod_{i=1}^{4} p^{T_i}(\vect{x_i} | y_i)
  \]
  for $W = R,S$, where $p^{T_i}(\vect{x_i} | y_i)$ is the
  probability of state vector $\vect{x_i}$ assuming the root of $T_i$ is in
  state $y_i$. This implies
  \[
  \begin{split}
    \alpha
    p_{\vect{\tilde{\gamma}_1}}^{\tilde{R}}(\vect{x_1},\ldots,\vect{x_4}) + (1-\alpha)
    p_{\vect{\tilde{\gamma}_2}}^{\tilde{R}}(&\vect{x_1},\ldots,\vect{x_4}) \\
    & = \sum_{\vect{y}} \, \left( \alpha p_{\vect{\gamma_1}}^R(\vect{y}) + (1-\alpha)
    p_{\vect{\gamma_2}}^R(\vect{y})  \right) \prod_{i=1}^{4} p^{T_i}(\vect{x_i} |
    y_i) \\
    & = p_{\vect{\tilde{\gamma}_3}}^{\tilde{S}}(\vect{x_1},\ldots,\vect{x_4})
  \end{split}
  \]
  where the $\vect{\tilde{\gamma}_j}$ are simply the $\vect{\gamma_j}$ along with
  the branch lengths of the $T_i$.

\end{proof}

For completeness we also record when a two class phylogenetic mixture
on a tree can mimic a tree of the same topology under the CFN model.
\begin{prop}
  If a two class phylogenetic mixture on a tree mimics a tree of the
  same topology under the binary symmetric model, then all branch lengths
  between the two sets must be the same with the possible exception of
  those for a quartet of adjacent edges sitting inside the tree.
\label{prop:same_big}
\end{prop}

\begin{proof}
  Assume a counter-example to Proposition~\ref{prop:same_big}: i.e.
  that there exists a tree $R$ with two branch length sets
  which differ by more than a quartet of adjacent edges
  but which mix to mimic a tree of the same topology $S$ under the binary
  symmetric model. Therefore, there exists a partitioning of $R$ into
  subtrees $A$, $B$, and $C$ meeting at a node such that
  there is an edge in each of $A$ and $B$ which differs in terms of
  branch length between the two sets. Note that if two
  branch length sets differ on a nontrivial rooted tree, then by
  induction one can find an induced rooted subtree of size two which
  differs in terms of branch length between the two branch length sets.
  Therefore there must be an induced rooted
  subtree of size two in each of $A$ and $B$ which differs in terms of
  branch length between the
  two branch length sets.  Number the taxa thus chosen from $A$ with 1 and
  2, and the taxa chosen from $B$ with 3 and 4.  Label an arbitrary taxon
  from $C$ with 5.
  Now consider the induced 5-taxon tree induced by restricting the taxon
  set to 1 through 5. Label the edges as in
  Figure~\ref{fig:fiveleaf}, and assign (induced) $k_i$'s as before.

  From the above we can assume (perhaps after renumbering) that $k_1
  \neq 1$ and $k_3 \neq 1$. By restricting $R$ to the taxon set $1234$ we
  have by Lemma~\ref{lem:same_top} that $k_1 = k_2^{-1}$ and $k_3 =
  k_4$ (perhaps after renumbering.)
  Because $k_3 \neq 1$, we have $k_3 \neq k_4^{-1}$.
  Thus using Lemma~\ref{lem:same_top}, by
  restricting to $1534$ we have $k_1 k_6 = k_5$ and by restricting to
  $2534$ we have $k_2 k_6 = k_5$.
  Therefore $k_1 = k_2 = 1$, which is a contradiction.
\end{proof}

\section{Conclusion}

We have presented a number of new results which help to
clarify when non-identifiable phylogenetic mixtures may pose a problem for
reconstruction. However, the message isn't completely straightforward.
The first section shows that the space of site-pattern frequency
vectors for
phylogenetic mixtures on many quartet trees contains a relatively large non-identifiable
region.  Furthermore, this non-identifiable region under the CFN model contains
site-pattern frequencies for resolved trees with substantial internal
branch lengths. Yet, these spaces were constructed using specific trees of
extreme branch lengths, raising the question of whether corresponding
results hold for more reasonable parameter regimes and ``random'' sets
of trees which one might find from data. Furthermore, we wonder if it is possible to
find simple $H$-descriptions of the phylogenetic mixture polytope for
larger star trees.

On the other hand, the second section shows generally that
phylogenetic mixtures on
just two trees may not pose so much of a problem. In particular, our results
make progress towards showing that clocklike two class phylogenetic mixtures 
may be identifiable under further assumptions. We also
show that pairs of trees under CFN rates-across-sites mixtures are
identifiable. Finally, we show that two class phylogenetic mixtures 
on a tree cannot ``change'' the topology too much.

In general, many interesting questions remain and we look forward to
seeing further progress in this field.

\subsubsection*{Acknowledgements}

The authors are grateful to Sebastien Roch for helpful discussions
on this work, and to the two anonymous referees for their many
helpful comments.

\end{spacing}

\bibliographystyle{plain}
\bibliography{structure}

\end{document}